\documentclass[prd,preprint,amsmath,nofootinbib,superscriptaddress]{revtex4}
\usepackage{graphicx}
\usepackage{bm}
\usepackage{epsfig}
\usepackage{color}
\usepackage{colordvi}
\usepackage{axodraw}

\newcommand{\beq}{\begin{equation}}
\newcommand{\eeq}{\end{equation}}
\newcommand{\bea}{\begin{eqnarray}}
\newcommand{\eea}{\end{eqnarray}}
\begin{document}



\title{\boldmath 
AdS/CFT duality for non-relativistic field theory\\
}

\author{Walter D. Goldberger}
\affiliation{Department of Physics, Yale University, New Haven, CT 06520}

\begin{abstract}

We formulate a correspondence between non-relativistic conformal field theories (NRCFTs) in $d-1$ spatial dimensions and gravitational theories in $AdS_{d+2}$ backgrounds with one compactified lightlike direction.   The breaking of the maximal $SO(2,d+1)$ symmetry of $AdS_{d+2}$ to the non-relativistic conformal group arises from boundary conditions on bulk fields, without the need to introduce non-vacuum sources of energy-momentum.      As a check of the proposal, we use the gravitational theory to reproduce the NRCFT state-operator correspondence between scaling dimensions of primary operators and energy eigenstates of the non-relativistic system placed in an external harmonic potential.

 \end{abstract}

\maketitle

\section{Introduction}

Non-relativistic quantum systems with two-body bound state energies tuned close to threshold seem to describe a variety of physical problems (see~\cite{Braaten:2004rn} for a review).    This tuning may be accidental, as it is in the effective theories that describe the few-body strong interactions of nucleons at low momentum transfers, or arranged by experimental manipulation, as with cold atoms placed in an external confining trap~\cite{corev}.   As the binding energy approaches threshold, the scattering length diverges and the precise nature of the two-body interactions becomes irrelevant.   In this limit, the dynamics can be formulated in terms of local non-relativistic scale invariant field theories (NRCFTs), in which the usual Galilean invariance of the interactions is enhanced to a non-relativistic conformal symmetry known as Schrodinger symmetry\footnote{Schrodinger invariant quantum field theories were first discussed in refs.~\cite{Hagen:1972pd,niederer}.    In $d=4$ spacetime dimensions, an important example for both nuclear and atomic physics is the theory of fermions at unitarity, consisting of spin$-1/2$ particles interacting through a local four-fermi operator.   This theory was shown to satisfy the Ward identities of the Schrodinger group in~\cite{Mehen:1999nd}, and the renormalization group scaling of theories closed to this fixed point was first described in~\cite{Kaplan}.  For lower dimensional examples see ref.~\cite{Jackiw:1990mb}.   Formal aspects of Schrodinger invariant field theories are discussed in~\cite{Henkel:1993sg,Nishida:2007pj}}.

Because the scattering length diverges, the NRCFTs relevant to nature are strongly coupled and thus cannot be treated by the usual perturbative techniques.   While it has been possible to obtain analytical results in the few-body sector (for a review of methods see~\cite{corev,castin}), the many-body properties of these theories have not yielded to analytical methods.   Instead one must turn to numerical simulations in order to account for phenomena observed in the laboratory.

In the case of strongly coupled, relativistic (super) conformal field theories, it has been possible in some cases to use the $AdS$/CFT correspondence~\cite{maldacena,Gubser,Witten:1998qj} to recast the dynamics in terms of a dual gravitational theory that, in a suitable limit, becomes weakly coupled.    It is therefore natural to ask if a similar mapping can be obtained for strongly coupled non-relativistic conformal theories.   Such a gravitational description is likely to yield robust results only when both sides of the duality have enough supersymmetry, so it is perhaps not directly applicable to NRCFTs that can be realized in the laboratory.   Nevertheless, establishing a weakly coupled dual gravitational description may still be useful.   For example, it might shed light on aspects of the dynamics that are universal and thus independent of any underlying supersymmetry.

A first step towards establishing the correspondence between $\mbox{NRCFT}_d$ (with $d-1$ the number of spatial dimensions) and  $(d+2)$-dimensional gravity theory was taken recently by Son~\cite{Son:2008ye} and by Balasubramanian and McGreevey~\cite{Balasubramanian:2008dm}.   In these papers, the usual $AdS$/CFT dictionary~\cite{Gubser,Witten:1998qj} is adapted to derive NRCFT correlators from a gravitational theory propagating in a background with isometry group $Sch(d-1),$ the $d$-dimensional Schrodinger group.   In order to generate this background, it is necessary include an energy-momentum term in the Einstein equations that breaks the maximal symmetry of the vacuum down to  $Sch(d-1)$.   The resulting metric in~\cite{Son:2008ye,Balasubramanian:2008dm} describes a static cosmological model driven by a pressureless fluid, and by applying the rules of $AdS$/CFT to bulk fields in this geometry, it is possible to reproduce the boundary correlators of Schrodinger invariant theories.

Here we propose instead a version of the correspondence in which the gravity side lives in a (locally) $AdS_{d+2}$ background.      The breaking of the full $SO(2,d+1)$ isometry group is achieved by imposing boundary conditions on $AdS$ bulk fields that only preserve $Sch(d-1)$.   As in~\cite{Son:2008ye, Balasubramanian:2008dm}, our prescription yields Schrodinger invariant field theory correlators on the boundary.   

There may be some advantages to formulating the gravity/NRCFT correspondence directly in $AdS$, without introducing additional non-vacuum energy-momentum to break the full conformal symmetry.  For instance, it may be easier to describe properties of NRCFT whose dual gravity description depend on the global structure of the spacetime (e.g. the state-operator map~\cite{Witten:1998qj,Horowitz:1998bj}, phase transitions~\cite{Witten:1998zw}).   Some of these global properties will be discussed below.   In addition, $AdS$ backgrounds emerge naturally in supergravity.  This raises the possibility of constructing explicit supersymmetric examples of $AdS$/NRCFT by taking the non-relativistic limit, along the lines discussed below, of standard $AdS$/CFT duality.     These supersymmetric examples  would allow one to verify some aspects of the approach discussed in this paper in a theoretically controlled setting (supersymmetric extensions of the Schrodinger algebra have been studied in~\cite{ss}).

In sec.~\ref{sec:poin} we set up the basic dictionary between NRCFT correlators and $AdS$ bulk physics in the Poincare patch.   We use the embedding of $Sch(d-1)$ into $SO(2,d+1)$ discussed in~\cite{Burdet:1977qw,Duval:1984cj,Henkel:2003pu} (see also~\cite{Son:2008ye}) to identify modes of $AdS_{d+2}$ bulk fields of definite non-zero momentum along a compactified lightlike direction with boundary NRCFT primary operators.    This yields a prescription for computing boundary correlators similar to that in~\cite{Son:2008ye,Balasubramanian:2008dm}.  In sec.~\ref{sec:ham} we discuss global aspects of the correspondence.   In particular, we develop the gravitational description of global Hamiltonian time evolution, and use it to reproduce the state-operator correspondence of NRCFTs established in~\cite{Nishida:2007pj} (see also ref.~\cite{Mehen:2007dn} for a generalization).   We also use the global description to recover the bound $\Delta\geq (d-1)/2$ of scaling dimensions in unitary NRCFTs.     By considering asymptotically-$AdS$ black holes, it is possible to extend the methods discussed here to the case of finite temperature/density NRCFTs.    This analysis will be presented in a separate paper~\cite{gr}.

\section{Non-relativistic CFTs}

A non-relativistic CFT is defined as field theory in $d-1$ spatial dimensions that is invariant under the symmetry group $Sch(d-1)$.   In this section we review the necessary facts about $Sch(d-1)$ symmetry and its relation to the conformal group $SO(2,d+1)$  that will be needed in the construction of $AdS$ duals later on.

\subsection{Schrodinger invariance}

The relation between NRCFTs in $d-1$ spatial dimension and $AdS_{d+2}$ spacetime is closely connected to the relation between the Galilean conformal symmetry of $(d-1)$-dimensional space, $Sch(d-1)$ and the conformal group $SO(2,d+1)$ of $(d+1)$-dimensional Minkowski space.    This connection is well known~\cite{Burdet:1977qw,Duval:1984cj,Henkel:2003pu}, and it is perhaps best illustrated in the context of a simple example~\cite{Duval:1994pw,Duval:1994vh}.   Consider a massless free scalar propagating in $d+1$-dimensional Minkowski space,
\begin{equation}
\label{eq:free}
S={1\over 2}\int d^{d+1} x \partial_\mu \phi \partial^\mu\phi.
\end{equation}
This is trivially invariant under the conformal group $SO(2,d+1)$.   Choose lightlike coordinates $(t,\xi)$ in Minkowski space, with
\begin{eqnarray}
\label{eq:tandxi}
\nonumber
t &=& {x^0 + x^d\over \sqrt{2}}, \\
\xi &=& {x^0 - x^d\over \sqrt{2}}.
\end{eqnarray}
The metric takes the form (with the notation ${\vec x}\cdot {\vec y} = x^i y^i$, $i=1,\ldots,d-1$),
\begin{equation}
\label{eq:mink}
ds^2 = - 2 dtd\xi + d{\vec x}\cdot  d{\vec x}.
\end{equation}
Concentrating on a single mode of $\phi$ with definite momentum in the $\xi$ direction, $\phi = e^{-i\xi m} \psi(t,{\vec x}),$ the action reduces to a free non-relativistic field theory (after renormalizing $\psi$),
\begin{equation}
S\rightarrow  \int dt d{\vec x}\,  \psi^\dagger(t,{\vec x}) \left(i\partial_t + {1\over 2 m}\nabla^2\right)\psi(t,{\vec x}),
\end{equation}
which is invariant under the group of non-relativistic conformal transformations, generated by  the usual Galilean generators $(H,P^i,K^i=\mbox{boosts},M^{ij}=\mbox{rotations}$), together with a dilatation $D$, a Galilean special conformal generator $C$, and a central charge $N$ corresponding to the total particle number.   For example, under dilatations, $D: (t,{\vec x})\rightarrow (\lambda^2 t,\lambda {\vec x})$ for some $\lambda$.   The complete algebra is given below in Eqs.~(\ref{eq:sch1})-(\ref{eq:sch3}).   

This example shows that $Sch(d-1)$, the symmetry group of a free non-relativistic field theory (or more generally of a non-relativistic CFT), can be viewed as the subgroup of $SO(2,d+1)$ that does not mix modes with different momentum $N$ along the null direction $\xi$.  In other words, it is the subgroup of $SO(2,d+1)$ that leaves a fixed lightlike momentum vector invariant~\cite{Burdet:1977qw}.

In this paper, we will take this as the definition of $Sch(d-1)$.    Let the invariant lightlike vector be
\begin{equation}
N={P^0 - P^d\over 2}.
\end{equation} 
Then the $Sch(d-1)$ algebra consists of $SO(2,d+1)$ generators that commute with $N$.     The algebra of the conformal group is spanned by the Poincare generators $(P^\mu,M^{\mu\nu})$ along with dilatations $\tilde D$ and special conformal transformations $K^\mu$.    It is given by
\begin{eqnarray}
\nonumber
[M_{\mu\nu},P_\rho] = -i(\eta_{\mu\rho} P_\nu -\eta_{\nu\rho} P_\mu),& [M_{\mu\nu},K_\rho]=-i(\eta_{\mu\rho} K_\nu -\eta_{\nu\rho}K_\mu),
\end{eqnarray}
\begin{equation}
[M_{\mu\nu}, M_{\rho\sigma}]=-i\eta_{\mu\rho} M_{\nu\sigma}+\mbox{perms.},
\end{equation}
\begin{eqnarray}
[{\tilde D},M_{\mu\nu}]=0 & [{\tilde D},K_\mu]=i K_\mu & [{\tilde D},P_\mu]=-i P_\mu,
\end{eqnarray}
\begin{equation}
[P_\mu,K_\nu]=2 i M_{\mu\nu}- 2 i \eta_{\mu\nu} {\tilde D}.
\end{equation}
The subalgebra that commutes with $N$ consists of the generators
\begin{eqnarray}
H={1\over 2} (P^0+P^d), & C = {1\over 2}(K^0-K^d), & {D} = M_{0d} +{\tilde D},
\end{eqnarray}
as well as  spatial translations $P^i$,  boosts $K^i=M^{0i}-M^{di}$, and spatial rotations $M^{ij}$.   These generators satisfy the (extended) $Sch(d-1)$ algebra.   whose non-zero commutators are
\begin{eqnarray}
\label{eq:sch1}
\nonumber
[M_{ij},P_k] = -i(\delta_{ik} P_j -\delta_{jk} P_i),& [M_{ij},K_k]=-i(\delta_{ik} K_j -\delta_{jk}K_i),
\end{eqnarray}
\begin{equation}
[M_{ij}, M_{rs}]=-i\delta_{ir} M_{js}+\mbox{perms.},
\end{equation}
\begin{eqnarray}
\label{eq:sch2}
[D,P^i] = -  i P^i ,& [D,K^i] =  i K^i, & [C,P^i]= i K^i,
\end{eqnarray}
\begin{eqnarray}
\label{eq:sch3}
[D,H]= - 2 i H , & [D,C]= 2i C,  & [H,C]= i D.
\end{eqnarray}
and 
\begin{equation}
[P^i,K^j]= - i N\delta^{ij}
\end{equation}
Note in particular that the generators $H,C,D$ form an $SL(2,{\bf R})$ subalgebra.    The eigenstates of the operator $N$ can be interpreted as states of definite particle number.   

\subsection{State-operator correspondence}
\label{sec:so}

In relativistic CFTs, there is a one-to-one map between local operators ${\cal O}(x)$ with definite scaling dimension $\Delta$ and the eigenstates of the conformal Hamiltonian ${\cal H}=(P^0+K^0)/2$~\cite{Luscher:1974ez}.   Roughly speaking, given an operator ${\cal O}(x)$, the state  $|{\cal O}\rangle = \lim_{x\rightarrow 0}{\cal O}(x)|0\rangle$ is an energy eigenstate, ${\cal H}|{\cal O}\rangle = \Delta |{\cal O}\rangle$.   

Nishida and Son~\cite{Nishida:2007pj} established an analog of the state-operator correspondence for NRCFTs, which we now review.  Let $|0\rangle$ be the vacuum state of $H$ and ${\cal O}(t,{\vec x})$ an NRCFT primary operator, which by definition satisfies
\begin{eqnarray}
\label{eq:prim}
[K_i,{\cal O}(0)]=0, & [C,{\cal O}(0)]=0, & [D,{\cal O}(0)]=-i\Delta {\cal O}(0).
\end{eqnarray}
Then it follows from the $Sch(d-1)$ algebra that the state 
\begin{equation}
|{\cal O}\rangle= e^{-{H/\omega}} {\cal O}^\dagger(0)|0\rangle,
\end{equation}
is an eigenstate of the deformed Hamiltonian $H_o=H+\omega^2 C$, where $\omega$ is a constant with units of frequency,
\begin{equation}
\label{eq:eigen}
H_o |{\cal O}\rangle = \omega \Delta |{\cal O}\rangle.
\end{equation}

The deformed Hamiltonian has a useful physical interpretation, which can be exhibited by noting that in a NRCFT, the special conformal generator $C$ can be realized in terms of the fundamental fields $\psi(t,{\vec x})$ as the operator
\begin{equation}
C={1\over 2} \int d{\vec x} \, \, \psi(t,{\vec x})^\dagger {\vec x}^2  \psi(t,{\vec x}).
\end{equation}
Thus the deformed Hamiltonian $H_o$ corresponds to placing the original NRCFT, with internal Hamiltonian $H$, in a harmonic external potential (setting to unity the mass of the particle excitations associated with the field operator  $\psi(t,{\vec x})$).     The state-operator map is therefore the statement that the eigenstates of the NRCFT in the harmonic trap are in one-to-one correspondence with the primary operators of the conformal theory, where the energy eigenvalues are the scaling dimensions\footnote{In addition to the correspondence between eigenstates of $H_o$ and the operator algebra in the free-space theory introduced in~\cite{Nishida:2007pj}, there is an additional correspondence between $H_o$ eigenstates and zero-energy homogeneous $N$-particle eigenfunctions of $H$~\cite{werner}.    As shown in ref.~\cite{Mehen:2007dn}, these, and many other such mappings, are related by inner automorphisms of the $SL(2,{\bf R})$ subalgebra of the Schrodinger group.}.   The $SL(2,{\bf R})$ subalgebra of $Sch(d-1)$ implies that above each state $|{\cal O}\rangle$, with ${\cal O}$ primary, there is a tower of $H_o$ eigenstates  with level spacing $2\omega$.    Different states in this tower are connected by the raising/lowering operators
\begin{equation}
\omega L_\pm =  H -\omega^2 C \mp i\omega  D,
\end{equation} 
which satisfy the commutation relations $[H_o,L_\pm]=\pm 2\omega L_\pm$ as a consequence of the $Sch(d-1)$ algebra.   In addition, the descendant operators ${\cal O}_{i_1\cdots i_\ell}(0)=\partial_{i_1}\cdots\partial_{i_\ell} {\cal O}(0)-\mbox{traces}$ generate $SL(2,{\bf R})$ towers above a base state with energy $\Delta+\ell$ in units of $\omega$.

The relation between time evolution generated by $H_o$ and the presence of a harmonic potential has a geometric interpretation, as pointed out in~\cite{Duval:1994vh}.   Following that reference, introduce the following coordinates on $(d+1)$-dimensional flat space:
\begin{eqnarray}
\nonumber
\omega t_o &=& \tan^{-1} \omega t,\\
\nonumber
\xi_o &=& \xi -  {1\over 2} \omega {\vec x}^2 {\omega t\over 1+ \omega^2 t^2},\\
{\vec x}_o &=& {{\vec x}\over \sqrt{1+\omega^2 t^2}}.
\end{eqnarray}     
These new coordinates have simple transformation properties under the action of $H_o$,
\begin{equation}
H_o:   (t_o,\xi_o,{\vec x}_o)\rightarrow (t_o+\tau, \xi_o,{\vec x}_o)
\end{equation}
for some constant $\tau$.  In these coordinates, the metric Eq.~(\ref{eq:mink}) becomes $ds^2 = \cos^2\omega t_o ds^2_o,$ with
\begin{equation}
\label{eq:mosc}
ds_o^2 = -\omega^2 {\vec x}_o^2 dt_o^2 - 2 dt_o d\xi_o + d{\vec x}_o^2.
\end{equation}
By covariantizing Eq.~(\ref{eq:free}), and adding a term proportional to $\int d^{d+1} x \sqrt{g} R \phi^2$ with a suitably chosen coefficient, the theory becomes invariant under arbitrary conformal transformations of $ds^2\rightarrow \Omega^2 ds^2$ of the background metric.      Making a transformation that converts Eq.~(\ref{eq:mink}) to the metric in Eq.~(\ref{eq:mosc}), and plugging in the momentum eigenmode $\phi=e^{-i m \xi_o}\psi(t_o,{\vec x}_o)$ into the conformally invariant free Lagrangian, one obtains (dropping the subscript $o$)
\begin{equation}
S\rightarrow \int dt d{\vec x} \, \psi^\dagger(t,{\vec x})\left(i\partial_t +{1\over 2 m}\nabla^2 -{1\over 2}m  \omega^2 {\vec x}^2\right)\psi(t,{\vec x}),
\end{equation}
which is the free NRCFT coupled to an external harmonic potential.

\section{AdS/CFT for non-relativistic field theories}

In this section we establish the correspondence between $\mbox{NRCFT}_d$ and gravity in $AdS_{d+2}$.   In order to generate a background with the requisite symmetry, one must break the full $SO(2,d+1)$ isometry group of $AdS$ down to a $Sch(d-1)$ subgroup.  This can be accomplished by selecting a preferred lightlike direction in $AdS$ generated by the orbits of the particle number operator $N$ in the algebra of $Sch(d-1)\subset SO(2,d+1)$.   The discreteness of the spectrum of $N$ suggests that the preferred lightlike direction should be compact.   

The Lagrangian form of the correspondence, suitable for computing correlation functions is developed in below in sec.~\ref{sec:poin}.    The procedure is the standard $AdS$/CFT prescription for Green's functions~\cite{Gubser,Witten:1998qj} (also employed in the approach of~\cite{Son:2008ye,Balasubramanian:2008dm}), with a modification of the asymptotic behavior of bulk fields near the $AdS$ boundary to account for the symmetry breaking $SO(2,d+1)\rightarrow Sch(d-1)$.   In sec.~\ref{sec:ham} we discuss the global Hamiltonian formulation.   There we show how the NRCFT analog of radial quantization (i.e., the state-operator map of ref.~\cite{Nishida:2007pj}) emerges in the $AdS$ picture.   

\subsection{Poincare coordinates:   correlation functions}
\label{sec:poin}

Consider a local NRCFT primary operator ${\cal O}_M$ (i.e., satisfying Eq.~(\ref{eq:prim})) with particle number $M$ and scaling dimension $\Delta$,
\begin{eqnarray}
[N,{\cal O}_M]=-i M {\cal O}_N, & [D,{\cal O}_M(0)]=-i\Delta {\cal O}_M(0).
\end{eqnarray}
The generating function of ${\cal O}_M(t,{\vec x})$ correlators is
\begin{equation}
\label{eq:genf}
Z_{NRCFT}[\phi]=\left\langle T \exp\left[i\int dt d{\vec x}\,  \phi(t,{\vec x}) {\cal O}_M(t,{\vec x})+\mbox{h.c.}\right]\right\rangle.
\end{equation}
where the expectation value is taken in the vacuum of the generator $H$.   We would like to establish a dictionary between correlation functions in  the $\mbox{NRCFT}_d$ and quantities in $AdS_{d+2}$.

To set up the correspondence, it is convenient to use coordinate on $AdS_{d+2}$ that point along the orbits of the generators $H$, corresponding to time evolution in the NRCFT (in the absence of external fields), and particle number $N$.    Because both these operators are linear combinations of translations $P^\mu$, we work in the Poincare coordinates  $X=(z,x^\mu)$ on $AdS$ which transform as $P^\mu: (x^\mu,z)\rightarrow (x^\mu+a^\mu,z)$.     The metric in these coordinates is
\begin{equation}
\label{eq:ppatch}
ds^2 = {\ell^2\over z^2} \left(-2 dt d\xi  + d{\vec x}\cdot  d{\vec x} + dz^2 \right)
\end{equation}
with $\ell$ the $AdS$ radius.    The null coordinates $(t,\xi)$ are given in terms of Poincare coordinates by Eq.~(\ref{eq:tandxi}), and simply shift under the action of $H, N$:
\begin{eqnarray}
H:(t,\xi)\rightarrow (t+\Delta t,\xi),& N:(t,\xi)\rightarrow (t,\xi+\Delta\xi).
\end{eqnarray}
Because  particle number $N=i\partial_\xi$ is discrete, we assume that the null direction parameterized by $\xi$ is compactified on a circle.    This is achieved by identifying $\xi\sim \xi+\lambda$ for some constant $\lambda.$   There is no invariant definition of radius for lightlike compactification, and indeed the compactification scale does not enter any of our results.   Note that the metric on the boundary, $z\rightarrow 0$, becomes the flat metric Eq.~(\ref{eq:mink}) up to an overall (divergent) factor.

The correspondence for correlation functions follows from the usual $AdS$/CFT dictionary between operators ${\cal O}$ and $AdS$ bulk fields,
\begin{equation}
\label{eq:adsnrcft}
Z_{NRCFT}[\phi] = Z_{AdS}[\phi].
\end{equation}
Here the right hand term is the effective action for a bulk field $\phi(X)$ in $AdS_{d+2}$ subject to the following asymptotics on the $AdS$ boundary at $z\rightarrow 0$,
\begin{equation}
\label{eq:asym}
\phi(X)\rightarrow z^{\Delta} e^{-i M\xi} \phi(t,{\vec x}).
\end{equation}
Assuming that the semiclassical approximation holds, one can replace $\ln Z_{AdS}[\phi]\simeq iS_c[\phi]$, where $S_c[\phi]$ is the classical gravity action in $AdS$.

To illustrates how this works, consider a free massive scalar field  $\phi$ propagating in $AdS_{d+2}$.  In the above coordinates the action is,
\begin{equation}
S= - \int d^{d+2} X \sqrt{g}\left[z^2\eta^{\mu\nu}\partial_\mu\phi^\dagger \partial_\nu\phi +  z^2 |\partial_z\phi|^2+ m^2 |\phi|^2\right].
\end{equation}
We set out to find a solution of the equation of motion with the boundary asymptotics Eq.~(\ref{eq:asym}).     This must be of the form 
\begin{equation}
\phi(X)= e^{-i M\xi} \phi_M(t,{\vec x},z).
\end{equation}
The equation of motion becomes
\begin{equation}
\label{eq:wavee}
-\partial_z^2 \phi_M +{d\over z} \partial_z \phi_M - (2 i M \partial_t  + {\nabla}^2)\phi_M + {{m}^2\over z^2} \phi_M=0.
\end{equation}
Note that near the boundary, the solutions behave as $\phi_M\sim z^{\Delta_\pm}$, with
\begin{equation}
\label{eq:andim}
\Delta_{\pm} = {d+1\over 2} \pm \sqrt{\left({d+1\over2}\right)^2 +m^2},
\end{equation}
which is the standard $AdS$/CFT relation between bulk mass and scaling dimension $\Delta$.  A similar relation between NRCFT scaling dimensions and bulk quantities was given in refs.~\cite{Son:2008ye,Balasubramanian:2008dm}.   However, our in our case the quantity $\Delta_{\pm}$ which governs the asymptotic behavior of the scalar field is independent of the eigenvalue $M$ of momentum in the $\xi$ direction\footnote{It is possible to modify Eq.~(\ref{eq:andim}) to include dependence on the momentum eigenvalue $M$.  This would arise, e.g., from  couplings to dynamical fields that get expectation values in the $\xi$ direction.    Such VEVs would shift the Lagrangian by the additional terms,
\begin{equation}
\label{eq:symb}
S_\xi = - \int d^{d+2} X \sqrt{g}\left[c_1 \phi^\dagger i\partial_\xi \phi + c_2 |\partial_\xi\phi|^2 +\cdots\right],
\end{equation}
which have the effect of replacing $m^2\rightarrow m^2 + c_1 M + c_2 M^2+\cdots$ in Eq.~(\ref{eq:andim}).}.  

Barring this discrepancy in the definition of $m^2$ appearing in Eq.~(\ref{eq:andim}), the wave equation Eq.~(\ref{eq:wavee}) for an eigenstate of $\xi$ momentum is identical to the corresponding field equations in~\cite{Son:2008ye,Balasubramanian:2008dm}.   This must be the case given that their background metric has ${Sch}(d-1)$ symmetry by construction.     In terms of Fourier modes along the $(t,{\vec x})$ directions, the solutions are as in, e.g.~\cite{Son:2008ye},
\begin{equation}
\label{eq:soln}
\phi_M \sim z^{d+1\over 2} K_\nu(Q z),
\end{equation}
with $\nu=\pm \sqrt{(d+1)^2/4 + m^2}$, and $Q=({\vec p}^2 - 2 M\omega)^{1/2}$.

Given the $z\rightarrow 0$ asymptotics of the Bessel function $K_\nu(z)$, the boundary behavior is indeed $\phi\sim z^{\Delta_{\pm}}$.   As discussed in~\cite{Klebanov:1999tb}, there is an ambiguity in the choice of root $\Delta_{\pm}$ for some values of the mass parameter $m$.   For $m^2>1-(d+1)^2/4$, finiteness of the (Euclidean) action fixes the choice $\Delta_+$, while for 
\begin{equation}
-\left({d+1\over 2}\right)^2 < m^2 < 1 - \left({d+1\over 2}\right)^2,
\end{equation}
both $\Delta_+$ and $\Delta_-$ are permissible.  The interpretation is that for each choice of root $\Delta$, the same $AdS$ theory gives rise to a different boundary CFT~\cite{Klebanov:1999tb}.    This interpretation carries over to NRCFTs, as discussed in~\cite{Son:2008ye}.   Note in particular that the smallest possible operator dimension is $\Delta = (d-1)/2,$ the dimension of a free non-relativistic field.  We will give a more careful derivation of this bound, along the lines of~\cite{BF}, in sec.~\ref{sec:ham} below from the point of view of Hamiltonian evolution.

Correlation functions of the primary operator ${\cal O}_M$ dual to $\phi$ can now be constructed using Eq.~(\ref{eq:adsnrcft}) and Eq.~(\ref{eq:soln}).   The calculation (see~\cite{Son:2008ye,Balasubramanian:2008dm}) is standard given the solutions Eq.~(\ref{eq:soln}) and will not be repeated here. For example, the two-point function, up to some normalization constant $c$, is given by
\begin{equation}
\label{eq:cor}
\langle T{\cal O}_M(t,{\vec x}) {\cal O}_M^\dagger(0,0)\rangle= c \theta(t) t^{-\Delta} \exp\left({i M \,{  {\vec x}^2\over 2t}}\right),
\end{equation}
where $\Delta$ is one of the roots in Eq.~(\ref{eq:andim}).   This is consistent with the results of~\cite{Henkel:1993sg, Nishida:2007pj} for the form of the two-point correlator of a primary operator ${\cal O}_M$ in an NRCFT.   Comparing Eq.~(\ref{eq:cor}) with the propagator of a free particle of mass $M$ in $d-1$ space dimensions
\begin{equation}
\langle {\vec x}| e^{-i H t}|{\vec x}=0\rangle = \left({M\over 2\pi i}\right)^{d-1\over 2}    \theta(t) t^{-{(d-1)\over 2}}\exp\left({i M \,{  {\vec x}^2\over 2t}}\right),
\end{equation}
suggests that that the $M$-particle bound state interpolated by ${\cal O}^\dagger_M$ can be interpreted as  a free particle of mass $M$ moving in a fractional $2\Delta$-dimensional space\footnote{Accordingly, the result in Eq.~(\ref{eq:eigen}) can be viewed as the ground state energy of a $2\Delta$-dimensional harmonic oscillator.}.   Connected correlators with more operator insertions encode scattering amplitudes.   These amplitudes are parameterized by the various coupling constants (e.g, gravitational couplings) in the $AdS$ bulk and thus scale roughly like powers of $G_N\ell^{-d}$, with $G_N$ the $(d+2)$-dimensional Newton constant.   If the $AdS$ radius $\ell$ is large in Planck units, these correlation functions can be computed from the usual $AdS$/CFT bulk-to-boundary diagrams, with boundary conditions as in Eq.~(\ref{eq:asym}) for each operator insertion.

\subsection{State-operator correspondence}
\label{sec:ham}

In relativistic $\mbox{CFT}_{d+1}$s, time evolution generated by the conformal Hamiltonian ${\cal H}$, whose eigenstates are the primary operators $|{\cal O}\rangle$, is equivalent to canonical quantization on the space ${\bf R}\times {\bf S}^d$~\cite{Luscher:1974ez}.    The scale introduced by the radius of the sphere ${\bf S}^d$ has the effect of generating a finite gap in the spectrum of CFT scaling dimensions.    ${\bf R}\times {\bf S}^d$ is also the global boundary of $AdS_{d+2}$, where ${\bf R}$ is spanned by $AdS$ global time $\tau$.   Given the correspondence between CFT operators and $AdS$ bulk fields, this implies a map between $AdS$ field eigenmodes of ${\cal H}=i\partial_\tau$ and the states $|{\cal O}\rangle$ of the boundary CFT~\cite{Horowitz:1998bj,Witten:1998qj}.   The spectrum of $i\partial_\tau$ is quantized due to the presence of a gravitational potential well for modes in $AdS$, and therefore matches the discreteness of the CFT spectrum on  ${\bf R}\times {\bf S}^d$.

In this section, we discuss the $AdS$ interpretation of analogous global aspects of NRCFTs.   In particular, we develop the correspondence between the state-operator map introduced in~\cite{Nishida:2007pj} and the eigenmodes of bulk fields with respect to a suitably chosen global time variable.   From this analysis, we also recover the unitarity bound $\Delta\geq (d-1)/2$ on $\mbox{NRCFT}_d$ scaling dimensions.   The consistency of these results with those of sec.~\ref{sec:poin} (e.g., the relation  Eq.~(\ref{eq:andim}) between scaling dimensions and $AdS$ field masses) provides a check of the $AdS$/NRCFT dictionary.

\subsubsection{Coordinates}

We need to introduce a time coordinate on $AdS_{d+2}$ that points along orbits of the oscillator Hamiltonian $H_o=H+\omega^2 C$ discussed in sec.~\ref{sec:so}.    A suitable set of coordinates can be obtained by starting with the definition of $AdS_{d+2}$ as the quadric surface
\begin{equation}
\label{eq:quad}
-\ell^2 = -2 X^+ X^- - 2 Z^+ Z^- + X^i X^i \,\,\,\,\,\,(i=1,\ldots, d-1)
\end{equation}
embedded in  ${\bf R}^{d+3}$ with metric
\begin{equation}
\label{eq:um}
ds^2 = - 2 dX^+ dX^-  -2 dZ^+ dZ^- + dX^i dX^i.
\end{equation}
We parameterize a given point in $AdS$ in terms of coordinates $(t,\xi)$ generated by the orbits of  $H_o$ and $N$ respectively and a set of spatial coordinates.   A given point in $AdS_{d+2}$ can be written as
($0\leq \omega t <2 \pi,$ $-\infty<\xi<\infty,$  $x^{\pm}>0$)
\begin{equation}
\label{eq:sdef}
\left(\begin{array}{c}
X^i\\
\omega Z^-\\
X^+\\
Z^+\\
\omega X^-\\
\end{array}\right) =  e^{-i t H_o} e^{-i\xi N} 
\left(\begin{array}{c}
x^i\\
0\\
x^+\\
0\\
\omega x^-\end{array}\right),
\end{equation}
where  $x^i, x^{\pm}$ satisfy the constraint (seting $\ell=1$ from now on),
\begin{equation}
\label{eq:const}
x^+ x^- = {1\over 2} (1 +r^2),
\end{equation}
($r^2=x^i x^i$).   Using the definition of $H_o$ and $N$ in terms of $SO(2,d+1)$ generators, we get 
\begin{eqnarray}
\nonumber
X^i &=&  \, \, \, \, x^i,\\
\nonumber
\omega Z^- &=& -x^+ \sin\omega t -\xi \omega x^-\cos\omega t,\\
\nonumber
X^+&=& \, \, \, \, x^+ \cos\omega t -\xi \omega x^-\sin{\omega t},\\
\nonumber
Z^{+} &=& -\omega x^- \sin\omega t, \\
\omega X^- &=&  \, \, \, \,\omega x^-\cos\omega t,
\end{eqnarray}
for the relation between the embedding coordinates and the coordinates $(t,\xi,x^\pm,x^i)$ adapted to the NRCFT.   The coordinates $(t,\xi,x^{\pm},x^i)$ with the above ranges cover the $AdS$ hyperboloid once.    In order to remove closed timelike curves, we unwrap the time coordinate $t$ by letting it go over  $-\infty<t<\infty$.    As in sec.~\ref{sec:poin}, the discreteness of the spectrum of $N$ suggests that the direction along the coordinate $\xi$ be compact.   This is achieved by identifying points $\xi\sim \xi + \lambda,$ for some constant $\lambda$.    The ambient metric Eq.~(\ref{eq:um}) in these coordinates is 
\begin{equation}
\label{eq:met}
ds^2 = -2 \omega^2 x^-(x^- d\xi + x^+ dt) dt - 2 dx^+ dx^- + dx^i dx^i,
\end{equation}
and the $AdS$ metric is the induced metric on points that satisfy the constraint Eq.~(\ref{eq:const}).    On surfaces of constant $(t,\xi)$, the metric is $ds^2 = - 2 dx^+ dx^- + dx^i dx^i$, and thus the coordinates $(x^\pm,x^i)$ subject to Eq.~(\ref{eq:const}) span a copy of $d$-dimensional hyperbolic space ${\bf H}^d$.   Eqs.~(\ref{eq:sdef}), (\ref{eq:met}) describe a locally $AdS$ spacetime with global topology ${{\bf S}^1}_\xi\times {\bf R}^{d+1}.$

To study the boundary (where the NRCFT lives), recall that in $AdS$ this consists of the re-scaled points  $(X^\pm,Z^\pm,X^i) \rightarrow  s (X^\pm,Z^\pm,X^i) $,  $s\rightarrow\infty$ of the quadric Eq.~(\ref{eq:quad}).    These points satisfy
\begin{equation}
0= 2 X^+ X^- + 2 Z^+ Z^- + X^i X^i,
\end{equation}
with the identification $(X^\pm,Z^\pm,X^i) \sim  s (X^\pm,Z^\pm,X^i) $ under (now finite) scaling.   On the coordinates $(t,\xi,x^\pm,x^i)$, this re-scaling gives
\begin{equation}
(t,\xi,x^\pm,x^i)\rightarrow (t,\xi, s x^\pm, s x^i).
\end{equation}
Thus the boundary consists of points spanned by $(t,\xi)$ as well as $(x^\pm,x^i)$, with the equivalence $(x^\pm,x^i)\sim s (x^\pm,x^i)$ under re-scalings, and subject to the constraint
\begin{equation}
\label{eq:bcon}
2 x^+ x^- = r^2.
\end{equation}
The set of points satisfying the relation consists of an ${\bf S}^{d-1}$, which is the boundary at infinity of hyperbolic space ${\bf H}^d$ as described by Eq.~(\ref{eq:const}).   Thus the boundary of the spacetime defined by Eq.~(\ref{eq:sdef}) is conformal to the space ${\bf R}_t\times {{\bf S}^1}_\xi\times {\bf S}^{d-1}.$

For example, using the equivalence under scaling, one can set, e.g., $x^-=1/2$, and thus $x^+=r^2$ by Eq.~(\ref{eq:bcon}).   The metric on the boundary becomes (up to an overall factor of $s^2\rightarrow \infty$)
\begin{equation}
ds_{bd}^2 = -\omega^2 r^2 dt^2 - {1\over 2}\omega^2 dt d\xi + dx^i dx^i.
\end{equation}
Up to normalization of $\xi$, this is exactly the metric in Eq.~(\ref{eq:mosc}). From the discussion in sec.~\ref{sec:so}, this indicates that the boundary NRCFT lives in a background harmonic confining trap.  Because Eq.~(\ref{eq:mosc}) is conformally flat, this result is consistent with the boundary conformal structure of $AdS$.    Alternatively, we can choose to set $x^+=1/2$ on the boundary by rescaling.     In that case, $x^-= r^2 $ and the boundary metric becomes
\begin{equation}
ds^2_{bd} = -\omega^2 r^2 dt^2  -2 \omega^2 r^4  d \xi dt + dx^i dx^i.
\end{equation}
This metric is conformally equivalent to the metric in the patch $x^-=1/2$, as can be seen by performing an inversion $x^i\rightarrow x^i/2 r^2$ and pulling out a factor of $r^4$.   The full boundary is the union of the two patches $x^{\pm}=1/2$.    

As it stands $x^{\pm}$ are not independent variables.    A convenient set of of coordinates on the surface Eq.~(\ref{eq:const}) is $x^i$ together with $z=x^+/x^-$.   In terms of $x^\mu=(t,\xi,z,x^i)$ the metric is now,
\begin{equation}
\label{eq:fmet}
ds^2 = -\omega^2 (1+r^2) \left(dt^2 +  {dt d\xi\over z}\right) +  {1\over 4}(1+r^2){dz^2\over   z^2} + {dr^2\over 1+r^2} + r^2 d\Omega_{d-2}^2,
\end{equation}
with $0<z<\infty$.       The boundary of the spacetime can be reached by taking $r\rightarrow\infty$.   To put the resulting metric in the form of the oscillator metric Eq.~(\ref{eq:mosc}), one has to re-define $z=\rho^2$ and make the conformal re-scaling $ds^2\rightarrow \rho^2 ds^2.$   Hypersurfaces $\Sigma_t$ of constant $t$ are lightlike, with normal vector $k=\partial_\xi$.   The generators of $\Sigma_t$ are the null geodesics with tangent vector $k$, one passing through every fixed value of $(z,x^i)$. 

\subsubsection{Energy eigenstates}

Having constructed the global time coordinate $t$, we can now check that the spectrum of $H_o$ in the NRCFT matches the energy spectrum of $i\partial_t$ acting on $AdS$ fields.   For simplicity, we consider the energy eigenmodes of a free massive scalar field $\phi$ in $AdS_{d+2}$.   The Klein-Gordon equation in the coordinates $x^\mu=(t,\xi,z,x^i)$ is 
\begin{equation}
\label{eq:KG}
\left[{4\over 1+r^2}\left(-{ z \over \omega^2} \partial_\xi\partial_t + {z^2\over\omega^2}\partial^2_\xi + z^2 \partial_z^2\right)+ {1\over r^{d-2} (1+r^2)}\partial_r(1+r^2)^2 r^{d-2}\partial_r + {1\over r^2}\nabla^2_{d-2}-m^2\right]\phi =0,
\end{equation}
plus possible terms from Eq.~(\ref{eq:symb}).    We will also need the norm for states quantized relative to (lightlike) hypersurfaces $\Sigma_t$ of constant $t$ ,
\begin{equation}
\label{eq:ip}
\langle \phi_1|\phi_2\rangle =  {i\over 2}\int_{\Sigma_t}  d\Sigma^\mu\,  \phi^\dagger_1 \stackrel{\leftrightarrow}{\partial_\mu} \phi_2 = {i\over 2}\int_{\Sigma_t} d\xi \, {dz\over 2 z} \, r^{d-2} dr\,  d\Omega_{d-2}  \, \phi^\dagger_1 \stackrel{\leftrightarrow}{\partial_\xi} \phi_2.
\end{equation}
In the second equality, we have used the definition $d\Sigma^\mu = k^\mu \sqrt{h} d^{d+1} x$, with $h$ the determinant of the metric on a $d$-dimensional surface with $t,\xi=\mbox{constant}$ and $k^\mu= \delta^\mu_\xi$ is the normal vector to $\Sigma_t$ (this is the definition of $k^\mu$,  $d\Sigma^\mu$ used in formulating Stokes' theorem for integrals over regions bounded by null surfaces).    The inner product is $t$-independent as long as the flux of the current  $J_\mu = i\phi^\dagger_1 \stackrel{\leftrightarrow}{\partial}_\mu \phi_2$ through the boundary at $r\rightarrow\infty$ vanishes,
\begin{equation}
\label{eq:flux}
{\cal F} = \left. r^{d+2} J_r\right|_{r\rightarrow\infty}\rightarrow 0.
\end{equation}

Plugging in the ansatz
\begin{equation}
\phi = e^{-i E t} e^{-i M\xi} Z(z) R(r) Y_\ell(\Omega),
\end{equation}
with $Y_\ell(\Omega)$ a spherical harmonic on ${\bf S}^{d-2}$, Eq.~(\ref{eq:KG}) factorizes into
\begin{equation}
\label{eq:rad}
{1\over r^{d-2}(1+r^2)} {d\over d r} (1+r^2)^2 r^{d-2} {d \over dr} R +\left({L^2\over r^2} + {m}^2 -{\lambda\over 1+r^2}\right) R =0,
\end{equation}
and
\begin{equation}
\label{eq:z}
{d^2 Z\over dz^2} +\left(-{M^2\over \omega^2}+ {M E\over\omega^2 z} -{\lambda\over 4 z^2}\right) Z =0.
\end{equation}
The effect of terms such as in Eq.~(\ref{eq:symb}) is to shift the bulk mass $m$ by some polynomial in $M$.  
Here $\lambda$ is a constant yet to be determined, and $L^2=\ell (\ell+d-3)$ is minus the eigenvalue of $\nabla^2_{d-2}$ acting on $Y_\ell(\Omega)$.   The solution of Eq.~(\ref{eq:rad}) that is regular at $r=0$ is the hypergeometric function
\begin{equation}
R(r)= r^\ell (1+r^2)^{-\Delta/2-\ell/2} {}_2 F_1\left(a,b;c;{r^2\over 1+ r^2}\right),
\end{equation}
with
\begin{eqnarray}
\label{eq:hyperdefs}
\nonumber
a &=& {1\over 2}(\ell+\Delta-1) + {1\over 2}\sqrt{1+\lambda},\\
\nonumber
b &=& {1\over 2}(\ell+\Delta-1) - {1\over 2}\sqrt{1+\lambda},\\
c &=& \ell  +{d-1\over 2}.
\end{eqnarray}
The parameter $\Delta$ is given by one of the roots in Eq.~(\ref{eq:andim}).    For Eq.~(\ref{eq:z}), the solution that is regular at $z\rightarrow 0$ can be written in terms of the confluent hypergeometric function
\begin{equation}
\label{eq:zhyper}
Z(z) =  z^{\mu+1/2} e^{-\omega z/{4 M}} {}_1 F_1\left(\mu-\nu+{1\over 2};2\mu+1;{\omega z\over 2 M}\right)
\end{equation}
with $\mu = {1\over 2}\sqrt{1+\lambda}$, $\nu = {E/2\omega}$.   

Near the boundary of spacetime, the radial function has the behavior
\begin{equation}
R(r\rightarrow\infty)\sim {\Gamma(a+b-c) \Gamma(c)\over\Gamma(b)\Gamma(a)}\,  r^{\Delta-d-1} + {\Gamma(c-a-b) \Gamma(c)\over\Gamma(c-a)\Gamma(c-b)}\, r^{-\Delta}.
\end{equation}
Normalizability with respect to Eq.~(\ref{eq:ip}) is only possible if $-b$ is a non-negative integer, in which case the term that scales like $r^{\Delta-d-1}$ vanishes.    The remaining term has finite norm as long as
\begin{equation}
\Delta\geq {d-1\over 2},
\end{equation}
and thus we recover the unitarity bound on NRCFT scaling dimensions from the gravity description.  This choice of parameters also implies that the flux Eq.~(\ref{eq:flux}) of the Klein-Gordon current vanishes at the boundary, ensuring that time evolution along $\Sigma_t$ hypersurfaces is unitary.   In addition, finiteness of the $z$ integral in Eq.~(\ref{eq:ip}) requires that $\mu-\nu+1/2$ in Eq.~(\ref{eq:zhyper}) must be a negative integer or zero.   We can now eliminate $\lambda$ to obtain a relation between $E$ and $\Delta$,
\begin{eqnarray}
\label{eq:result}
{E\over 2\omega} =  {1\over 2} (\Delta +\ell-1) +n,\,\,\,\,\,\,\,\,\,  (n=0,1,2,\cdots).
\end{eqnarray}

Up to a zero-point constant, this result reproduces all the features of the NR state-operator correspondence developed in~\cite{Nishida:2007pj}.     A primary operator ${\cal O}(0)$ of the boundary NRCFT is dual to the $\ell=0$ mode of the field $\phi$ and corresponds to the lowest level in an $SL(2,{\bf R})$ ladder of oscillator states with spacing $\Delta E= 2\omega$.   Higher multipoles of $\phi$  correspond to the states  ${\cal O}_{i_1\cdots i_\ell}=\partial_{i_1}\cdots \partial_{i_\ell}{\cal O}(0)-\mbox{traces},$ of scaling dimension $\Delta+\ell$.  Eq.~(\ref{eq:result}) confirms that the duality between NRCFT operators and $AdS$ bulk fields proposed in sec.~\ref{sec:poin} correctly reproduces the global aspects of Schrodinger invariant field theory.

\section{Conclusion}

In this paper we have formulated a correspondence between (semiclassical) gravity in $d+2$ dimensions and non-relativistic conformal field theory in $d-1$ spatial dimensions.   In this approach, the $Sch(d-1)$ symmetry of the NRCFT manifests itself on the gravity side as the residual symmetry of $AdS$ that remains after projecting onto eigenmodes along a fixed null direction.   It becomes possible to set up the correspondence directly in (locally) $AdS$ spacetimes, avoiding the need to introduce non-vacuum sources of energy-momentum.

Here we have set up the basic dictionary (similar to the usual $AdS$/CFT dictionary) and tested the consistency of the approach by showing that it reproduces certain properties of NRCFTs that can be derived from symmetry considerations alone (the form of two-point correlators and the state-operator correspondence).  Although we have only explicitly considered a toy gravity theory containing a single dynamical scalar, it  is straightforward to extend our analysis to more complicated gravity models.      In particular, to reproduce the operator spectrum of realistic NRCFTs it necessary to include additional fields, for example:    the bulk graviton $h_{\mu\nu}$ (related to the various components of $T_{\mu\nu}$ in the boundary theory), a Kaluza-Klein gauge field resulting from lightlike compactification (dual to the particle number current operator), bulk gauge fields related to the global symmetries of the boundary CFT (e.g., an $SU(2)$ gauge field dual to intrinsic spin on the CFT side).   In addition, fermionic bulk fields need to be introduced to describe both real world NRCFTs and supersymmetric examples.

It remains to be seen whether the $AdS$/NRCFT proposal presented here can be used to make predictions for the dynamics of strongly coupled NRCFTs probed in the laboratory.     As a first step in that direction, it is necessary to establish a version of the correspondence for NRCFTs at finite temperature and density.   It seems clear that this can be done by studying gravity duals propagating in asymptotically $AdS$ (charged) black hole backgrounds, compactified on a suitable chosen lightlike direction in order to generate the appropriate non-relativistic symmetries.   In that case it should be possible to apply the existing $AdS$/CFT technology for e.g., computing finite temperature/density response functions and hydrodynamic transport properties~\cite{PSS1,Birmingham:2001pj,SStHS,PSS2}.

\centerline{\bf Acknowledgements}
I thank Ira Rothstein for valuable discussions and for collaboration at early stages of this project.   This work is supported in part by DOE grant DE-FG-02-92ER40704, and by an OJI award from the DOE.   Some of the work presented here was carried out at the Kavli Institute for Theoretical Physics and is supported in part by NSF grant PHY05-1164.

\end{document}